A SEARCH FOR DYING PULSE TRAINS IN CYG X-1 USING *RXTE*

Joseph F. Dolan and Daria C. Auerswald

Department of Astronomy

San Diego State University

San Diego, CA 92182-1221

tejfd@sciences.sdsu.edu

E-mail: tejfd@sciences.sdsu.edu

# ABSTRACT

Dying pulse trains (DPTs) – pulses of radiation with decreasing intensity and decreasing intervals between them – are predicted by General Relativity to occur from material spiraling into an event horizon after detaching from the last stable orbit in an accretion disk around a black hole. Two events resembling DPTs were detected in 3 hours observation of Cyg X-1 in the far UV using the High Speed Photometer on the Hubble Space Telescope (Dolan 2001). We observed Cyg X-1, a leading candidate for a black hole, with the proportional counter array on RXTE to seek such events in the low-energy X-ray region. No dying pulse trains with a characteristic timescale between pulses of 1 - 40 ms were detected in 10 hours of observation during Cyg X-1's high luminosity state, low luminosity state, and transitions between states, although individual pulses are clearly detectable in data with 1 ms temporal resolution. The $2\sigma$ upper limit on the rate of DPT's in the X-ray region is less than half the rate reported by Dolan (2001) in the UV. These negative results are consistent with Cyg X-1 being an extreme Kerr black hole with a characteristic timescale between DPT pulses less than 1 ms.

Key Words: accretion, accretions disks – black hole physics – relativistic processes – X-rays: individual (Cyg X-1)

## 1. INTRODUCTION

A black hole is a point singularity surrounded by an event horizon that is predicted to exist by general relativity. For an uncharged Schwarzschild black hole having dimensionless angular momentum $a = cJ/GM^2 = 0$, where $M$ is the mass of the black hole and $J$ is its angular momentum, the event horizon lies at distance $R_S = 2GM/c^2 = 3m$ km, where $m$ is the mass of the black hole in units of solar mass (Misner, Thorne and Wheeler 1973). In geometric units, where $c = G = 1$, $0 \le a \le 1$. (Some authors adopt the convention that a is negative when the spin of a black hole in a binary system is retrograde to its orbit.) Stable Keplerian orbits around a Schwarzschild black hole exist for $r \ge 3\ R_S$; their orbital period is

$$P = 0.6m(r/3R_S)^{3/2} \text{ ms} \qquad [1]$$

(Sunyaev 1973). The innermost stable circular orbit closest to the event horizon, defining the inner edge of the accretion disk, is at $r = 3R_S$. The Keplerian orbital period at this distance is $P_O = 0.6m$ ms.

For $a > 0$ (a Kerr black hole), $P_O$ is is smaller than its value for a Schwarzschild black hole. For $a = 1$, the maximum value allowed, $P_O = 0.06m$ ms (Novikov & Thorne 1973).

Flare patches – self luminous clumps of emitting material whose enhanced or unusual radiative characteristics stand out above the mean flux of the system – are likely to exist in the accretion disk (Sunyaev 1973; Novikov & Thorne 1973). For lines of sight close to the plane of the accretion disk, such flare patches will appear to emit a pulse of radiation once each orbit around the black hole as viewed by a distant observer because of Doppler and abberational effects (Cunningham & Bardeen 1972). At the inner edge of the accretion disk the interval between pulses would be $P_O$. Stoeger (1980) reviewed the theoretical model of accretion disks and pointed out that a series of pulses would be observed from a flare patch detaching from the inner edge of an accretion disk and spiraling into the event horizon. The characteristic interval between pulses will be close to the orbital period of the emitting material and will decrease with time because the flare patch orbits the black hole more quickly as it approaches the event horizon. On the order of 10 pulses are expected to be visible during the infall; the interval between the pulses will gradually decrease to ~(0.6 – 0.9)$P_O$, depending on the number of pulses visible. Individual pulses are expected to have a FWHM ~ 10% of the interval between them, with the pulses originating closer to the event horizon being broader than those originating farther away (Cunningham & Bardeen 1972; de Felice, Nobili & Calvani 1974).

The pulsed fluxes observed from a flare patch falling into an event horizon will become progressively weaker because the flare patch's emission is redshifted toward infinite wavelength as it approaches the even horizon. Stoeger (1980) coined the phrase 'dying pulse train' (DPT) to refer to a series of pulses separated from each other by ever-decreasing intervals, with a characteristic timescale near $P_O$ and decreasing intensity with increasing pulse number. A DPT is a characteristic signature of a compact accreting object with an event horizon, i.e., a black hole.

Dolan (2001, hereafter D01) detected two possible DPTs in the Cygnus X-1/ HDE226868 binary system during 3 hours observing time in a 1400 – 3000 Å bandpass using the High Speed Photometer (HSP) on the Hubble Space Telescope (*HST*). Cyg X-1 is a leading candidate for a black hole, in the sense that a neutron star or white dwarf has been ruled out by previous observations, although these observations do not rule out the possibility that it may be a more exotic object (Bahcall, Lynn & Selipski 1990). For Cyg X-1, $m$ ~10 (cf. Caballero-Nieves et al. 2009 and the references therein), so the DPT pulses predicted by Eq. [1] for a Schwarzschild black hole would have FWHM ~ 1 ms. Further UV observations of the system were prevented by the removal of the HSP from *HST* in 1992 and the lack of any other satellite-borne instruments capable of making photometric observations in the UV with < 1ms time resolution.

We observed Cyg X-1 in the low-energy X-ray region using the proportional counter array on the Rossi X-Ray Timing Explorer (*RXTE*) in an attempt to confirm the phenomena reported by D01. If DPTs are detectable in the UV spectrum of the system, which is dominated by the O9.7 Iab primary star, then they should also be visible in the system's 2-15 keV spectrum, which is dominated by the accretion disk around the X-ray source. If no DPTs are detected in the X-ray region, and the upper limit on their frequency is significantly below that reported by D01, then the results reported by D01 may be spurious, or else DPTs may occur only when Cyg X-1's accretion disk attains a specific geometric structure that did not occur during the observations.

## 2. OBSERVATIONS AND DATA ANALYSIS

We searched for DPTs in the observations obtained in our regular monitoring program of the X-ray spectrum of Cyg X-1 from 1996 to 2006 using *RXTE* (Wilms et al. 2006). Photometric data obtained with the Propotional Counter Array (PCA) had a temporal resolution of 122 µs. Typical energy bandpasses extended from 2 – 15 keV, which we divided into two separate energy ranges having approximately equal counting rates. (In the 2 – 15 keV bandpass, these energy ranges were 2 – 6 and 6 – 15 keV.) Other bandpasses we observed incuded 1 - 5 / 5 - 13 keV and 1 - 9.5 / 9.5 -73 keV. The photons in each energy range were independent, in the sense that any photon detected in one energy range was not present in the other.

The 10 hours of observations we report here include 3.68 h when Cyg X-1 was in its high state (including 1.77 h obtained 2 d before a high to low state transition); 3.58 h in its low state; 1.67 h during a low state to high state failed transition; and

1.02 h during a failed low to high transition during the start of the drop back to the low state on 1999 Sept. 25.

The protocol adopted for searching for DPTs was the same as described in D01. We define a pulse as a statistically significant positive intensity variation having a full width at half maximum (FWHM) of a specified duration or shorter that is simultaneously present in the two independent energy ranges. Additional criteria used to discriminate between pulses and stochastic variability in the source are given in D01. Although a frequency of occurrence analysis can estimate how likely (or unlikely) it is that any intensity excursion is a stochastic variation in the source's flux, it can never identify any single event as a pulse because, given enough observing time, stochastic variability can mimic any characteristic of a photon pulse caused by a flare patch.

Because the X-ray flux from Cyg X-1 exhibits stochastic variability over timescales shorter than 1 s, the mean flux to which the significance level of a variation must be referred is the local mean flux of the source in a time interval including the variation. We estimated the local mean flux from the counting rate during a 100 ms interval prior to and a 100 ms interval immediately after the peak counting rate of the variation. The two intervals from which the mean counting rate was estimated included the underlying counts in the wings of any pulse, but excluded the peak count rate channels. If the variation is stochastic and not a pulse, then excluding the peak count rate region from the calculation of the local mean biased the mean to lower values. We therefore adopted a mean flux level 10% above that calculated to attempt to to correct for this bias.

We searched for pulses with FWHM between ~0.25 and ~10 ms. The FWHM duration of the event was estimated as the time at the center of the sample interval on the descending slope of the pulse containing half the counts of the peak sample interval (after subtracting the counts per sample in the continuous flux) minus the time at the center of the corresponding sample interval on the ascending slope.

## 3. RESULTS

### 3.1 Pulses

Intensity variations fulfilling all the criteria for pulses (D01) were detected from Cyg X-1 during all the luminosity states in which we observed it. These events are undoubtedly the same phenomena detected as 'powerful millisecond flares' by Gierlinski and Zdziarski (2003). A typical example of such an event is shown in Fig. 1, which occurred on1996 February 12 at 09:39:20 UT when Cyg X-1 was in the low state. The PCA counting rates in the 1 – 73 keV bandpass and in the 1 – 9.5 and 9.5 – 73 keV energy regions are binned with a 4.88 ms sample time. The bin numbers on the X-axis are the number of 4.88 ms time intervals since the (arbitrary) start of the data set. The event we identify as a pulse peaks at 105 counts in sample interval 4192. We assign a local continuous flux level of 40 counts per sample interval at the time of the event. The Poisson probability of observing a sample interval with 105 counts or more in a data set with a mean of 40 counts per sample interval is $10^{-23}$. There are 20,480 sample intervals in the 100 s long data set long data set, so the expectation value for the number of sample intervals with a count of 105 or larger is $1.5 \times 10^{-19}$. The data binned at 4.88 ms has 40 different possible starting phases of the 122 µs sample time observations. If we treat each of these different phases as independent trials and multiply by 40, the probability of the event being a stochastic variation is $10^{-17}$. (We emphasize, however, that the 40 data sets with different starting phases are not independent. The event appears at a similar level and with a similar shape in all 40 phases. Including this factor conservatively overestimates the probability of the event being a stochastic variation.) Multiplying again by 28.8, the number of separate 100 s duration observations of Cyg X-1 we examined at this epoch, and again by 7, the number of different binning times we examined for each 100 s observation (which again are not independent trials), the probability of this event occurring stochastically in any of the data we examined at this epoch is $< 10^{-14}$.

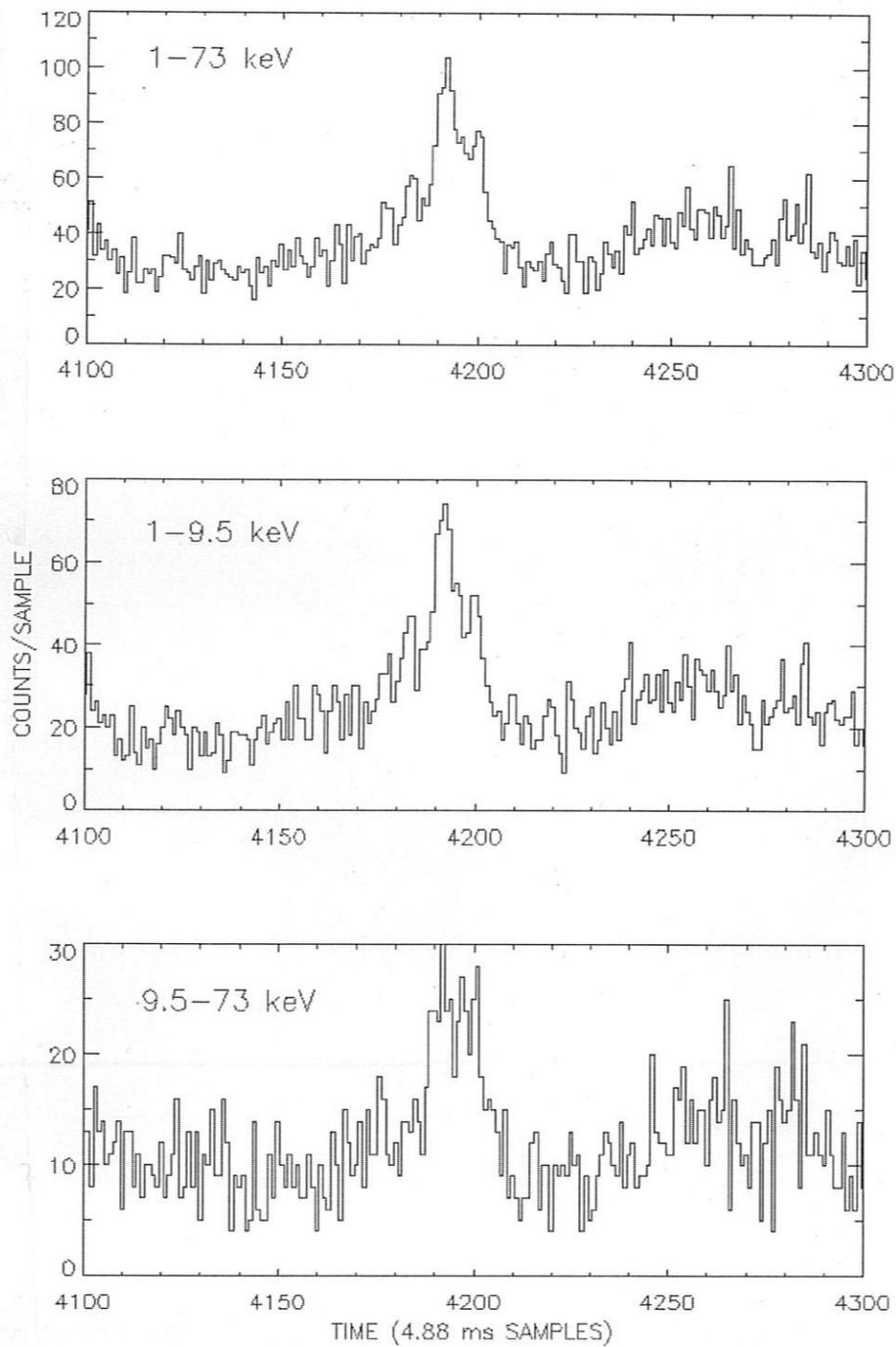

Figure 1. The PCA counting rate from RXTE on 1996 February 12, 09:39:20 UT at 4.88 ms sample time. The time is given as the number of sample intervals after the start of the data set. The event peaks in sample interval 4192.

Additional evidence exists for the identification of this event as a ms-burst of radiation from the source:

- The FWHM of the event, given a continuous flux level of 40 counts per sample interval, is 5 sample intervals. The number of detected photons in these 5 sample intervals, which we define as the fluence of the pulse, is 454 counts. The fluence expected in the same 5 intervals from the continuous flux is 200 counts. The Poisson probability of obtaining a sample interval with 454 counts in a distribution with an expectation value of 200 counts is $10^{-73}$.
- The event occurs at the same time and with a similar shape in both the 1 – 9.5 keV energy region and the 9.5 – 73 keV energy region (cf. Fig. 1).
- The event shows a smooth rise to and fall from maximum when examined at faster time resolution.
- The FWHM, $\Delta\tau$, of this pulse is 24 ± 4 ms, corresponding to a light travel time of $c\Delta\tau$ = 7300 ± 1200 km. This is a distance scale typical of the inner accretion disk around Cyg X-1.

The point of this example is not whether these intensity excursions are pulses or stochastic variations. Rather, it shows that if ms X-ray pulses as we define them in D01 are present in Cyg X-1, then we can detect them in our photometric observations.

**3.2 DPTs**

A series of 3 or more pulses having the characteristics defined in D01 would be a candidate dying pulse train. No DPTs were detected in the 10 hours of X-ray photometry we examined from Cyg X-1.

Candidate DPTs occurred in UV observations of Cyg X-1 (D01) with frequency 0.7 hr$^{-1}$ over 3 hours observing time. If the rate of candidate DPT events without regard to the luminosity state of the system is the same in the low-energy X-ray region as it is in the UV, the expectation value for detected DPT candidates in 10 hours would be 7. The Poisson probability of detecting 0 events for an expectation value of 7 is 9 x 10$^{-4}$, corresponding to a 3.3σ deviation in Gaussian statistics. The 2σ upper limit on the occurrence of DPTs in our observations is 2.8, corresponding to a frequency of occurrence over all observing time < 0.3 hr$^{-1}$.

## 4. DISCUSSION

If black holes exist, have event horizons, and exhibit pulses from flare patches, then dying pulse trains should appear in the photometry of the source (Stoeger 1980). Events having the characteristics we define as pulses are observed in the low-energy X-ray region of Cyg X-1, and these imply the existence of flare patches in the accretion disk. The lack of DPTs in our observations implies that physical conditions exist in the Cyg X-1 system that prevent their detection.

The upper limit on the frequency of DPTs (without regard to the luminosity state of the system) that we derive for low-energy X-rays is inconsistent with the frequency of candidate DPTs observed in the UV by D01. If the candidate DPTs in the UV are not spurious stochastic variability, then it may be possible that DPTs exist only when the accretion disk exhibits a specific structural geometry. If we did not survey the system during the X-ray luminosity state related to this geometry, then we would not have detected DPTs. (The two candidate UV DPTs did occur within 1.7 hr of each other, when the geometry of the disk presumably was the same for both.)

Models of the X-ray continuum emitted by an accretion disk around a black hole in a binary system, together with the values of its mass and the inclination of the system, can be used to provide an estimate of the dimensionless angular momentum of the black hole (Zhang et al. 1997; Brenneman and Reynolds 2006; Steiner et al. 2009). Applications of this method to systems in which the black hole is the primary consistently indicate that these objects are extreme Kerr black holes, i.e., have $a > 2/3$ (Zhang et al. 1997; Gierlinski et al. 2001; Shafee et al. 2006; Liu et al. 2008; Gou et al. 2009). There are systems in which the black hole is the secondary (like LMC X-3) where $a < 0.26$ (Zhang et al. 1997; McClintock et al. 2006; Davis et al. 2006). The extreme Kerr black hole primaries appear to have been formed in that state and not spun up gradually via accretion torques (McClintock et al. 2006).

Zhang et al. (1997) estimate the dimensionless angular momentum of Cyg X-1 using the continuum fitting model as $a \sim 0.75$, which would make it an extreme Kerr black hole. For $m \sim 10$, as in Cyg X-1, equation [1] gives the characteristic timescale between pulses in a DPT in an extreme Kerr black hole as < 1 ms. The FWHM of these pulses would then be < 0.1 ms. The parameter space we searched for DPTs was not well matched to either timescale. The low-energy flux of Cyg X-1 had a typical expectation value of < 2 detected counts in a 122 μs sample time in our RXTE data. If Cyg X-1 is an extreme Kerr black hole, DPTs would not cause significant enough increases in this counting rate to be detectable in our observations. Further, the FWHM of the each pulse would be less than the fastest sample time we used, lowering the statistical significance of any individual pulse (D01). Our negative result does not rule out the existence of DPTs from Cyg X-1 if it is an extreme black hole.

If DPTs are exhibited only when the accretion disk surrounding Cyg X-1 attains a specific geometric structure (which is currently not predictable from theory), then continued analysis of the source in low-energy X-rays is the only feasible method of attempting to detect them. Alternatively, if Cyg X-1 is an extreme Kerr black hole, the detection of DPTs must await the next generation of X-ray satellites, with microsecond time resolution photometry, as in RXTE, but larger detected counting rates, i.e., larger sensitive detecting areas.

**Acknowledgements.** Some of this work was carried out when JFD was at NASA's Goddard Space Flight Center. We thank Julia N. Holland for assisting in the analysis of the data, and an anonymous referee for pointing out the paper by Gierlinski and Zdziarski. We acknowledge support from the Research Experience for Undergraduates program at San Diego State University.